\documentclass{article}

\usepackage{arxiv}
\usepackage[utf8]{inputenc}

\usepackage[
    natbib=true,
    style=numeric,
    sorting=none
]{biblatex}
\usepackage{biblatex}
\usepackage[english]{babel}
\usepackage{csquotes}
\usepackage[T1]{fontenc}    
\usepackage{hyperref}
\hypersetup{
    colorlinks=true,
    linkcolor=red,
    filecolor=magenta,      
    urlcolor=blue,
    pdftitle={Overleaf Example},
    pdfpagemode=FullScreen,
    }
\usepackage{authblk}
\usepackage{booktabs}       
\usepackage{amsfonts}       
\usepackage{nicefrac}       
\usepackage{microtype}      
\usepackage{lipsum}		
\usepackage{amsmath}
\usepackage{graphicx}
\usepackage{doi}
\usepackage{graphicx}
\usepackage{dcolumn}
\usepackage{bm}
\usepackage{amssymb}
\newtheorem{theorem}{Theorem}

\title{Non-Null and Force-Free Electromagnetic Configurations in Kerr Geometry}

\author[1,2,\thanks{radhikari@troy.edu}]{Rakshak Adhikari } 
\author[2]{Govind Menon}
\author[1,3,4,5]{Mikhail V. Medvedev}
\affil[1]{Department of Physics, University of Kansas, Lawrence, KS 66045}
\affil[2]{Department of Chemistry and Physics\\ Troy University, Troy, AL 36082}
\affil[3]{Institute for Advanced Study, School for Natural Sciences, Princeton, NJ 08540}
\affil[4]{Department of Astrophysical Sciences, Princeton University, Princeton, NJ 08544}
\affil[5]{Laboratory for Nuclear Science, Massachusetts Institute of Technology, Cambridge, MA 02139}

\renewcommand{\shorttitle}{FFE in Kerr}

\begin{document}
\maketitle
\begin{abstract}
Force-Free Electrodynamics (FFE) is known to describe the highly magnetized plasma around pulsars and astrophysical black holes. The equations describing FFE are highly non-linear and the task of finding a well-defined analytical solution has been unyielding. However, FFE can also be understood in terms of 2-dimensional foliations of the ambient 4-dimensional spacetime. The study of the foliations can provide significant insights into the structure of the force-free fields and such foliations can be exploited to generate new null and non-null solutions. In this paper, we present several non-null solutions to the FFE equations in the Kerr spacetime obtained by applying the aforementioned methods of non-null foliations.
\end{abstract}


 \section{Introduction}
Many astrophysical systems, such as pulsars, magnetars, and rotating black holes, possess highly magnetized, plasma-filled magnetospheres where the energy stored in the field is significantly greater than the rest mass energy of the plasma.  Such systems are characterized by the vanishing Lorenz force, and their physics is governed by the equations of force-free electrodynamics (FFE). When plasma inertia can be ignored, FFE appears to be the leading theory for describing a variety of powerful phenomena, such as jets associated with pulsars and AGNs, as well as gamma-ray bursts.

Using FFE, Blandford and Znajek showed in their seminal paper \cite{Blandford:1977ds} that it is possible to directly extract electromagnetic energy from black hole spin for a slowly rotating black hole. Similarly, Goldreich and Julian showed \cite{GJ69} that a steady force-free magnetosphere exists outside a pulsar modeled as a conducting spherical star threaded on the surface by a poloidal magnetic field. The prevailing electromagnetic fields allow for the extraction of rotational energy via pulsar winds in the form of a Poynting flux. 

Despite decades of work, the efforts to generalize Blandford and Znajek's perturbative solution to a physically relevant exact solution have been intractable. By now, however, two classes of exact solutions in Kerr have emerged. A class of exact solutions in Kerr geometry valid at the event horizon was first given by Menon and Dermer \cite{MD07} nearly thirty years after the publication of \cite{Blandford:1977ds}. Recently, a generalization to this solution was obtained by Brennan et al. by employing the Newman-Penrose tetrads and assuming the current to be aligned to one of the null tetrads \cite{Brennan:2013kea}. Analytical work on force-free electrodynamics has historically been done in 3 + 1 formalism. However, recent works \cite{GT14} have shown that methods of exterior calculus can accentuate the rich geometric structure of the equations of FFE and make certain calculations simpler.

While exact solutions to FFE have been rare, higher-order perturbative solutions to the equations of FFE have continued to remain an active area of exploration \cite{Armas_20} \cite{Camilloni:2022kmx}. Considerable work has also been done in obtaining numerical solutions describing magnetospheres of both neutron stars and accreting black holes \cite{Mahlmann:2020nwe} \cite{Mahlmann:2020yxn} \cite{Qian:2018qpg} \cite{Mahlmann:2021yws} \cite{Pan:2018zsd} \cite{Petri:2021pxg} \cite{Carrasco:2019aas}. Observations of relativistic jets indicate a strong positive correlation between black hole spin and the power of the relativistic jets, thus supporting the rotational energy extraction mechanism \cite{Steiner:2012ap}. General Relativistic Magnetohydrodynamic (GRMHD) simulations \cite{Penna:2013rga} of rotating black holes have provided further support to the feasibility of the Blandford-Znajek mechanism as the source for relativistic jets coming from AGN.

The study of FFE and its observational effects may serve as a test for general relativity in the strong regime and recently there has been work in exploring FFE and the BZ mechanism in both modified gravity theories\cite{Dong:2021yss} as well as in metrics other than the Kerr metric \cite{Pei:2016kka} \cite{Konoplya:2021qll} \cite{Camilloni:2023wyn}. Because energy extraction takes place near the horizon, the Force-Free paradigm can be explored in the simplified metric describing the region near the horizon of an extremal Kerr black hole. Because of the simplified nature of this metric, many classes of exact solutions are possible\cite{GLS_16} \cite{Lupsasca:2014pfa} \cite{Cam_20} \cite{Compere:2015pja}. 

After a brief introduction to the methodology for the search for force-free solutions, we immediately focus on the exterior Kerr geometry and present all the previously known degenerate non-null solutions. Following this introduction and recapitulation, we present four different classes of exact non-null FFE solutions in Kerr geometry. Our solutions triple the catalog of previously known solutions. However, an exact solution describing energy extraction from Kerr black hole continues to remain elusive. It is our hope that the techniques presented here will serve as a backdrop for future investigations.

\section{Equations of Force-Free Electrodynamics}
While the solutions to the equations of Force-Free Electrodynamics presented in this paper are in the Kerr spacetime, the formalism derived in this section applies to all electrically neutral spacetimes. Given a fixed, electrically neutral background spacetime ${\cal M}$, electromagnetic field tensor $F$ satisfies
\begin{equation}
 d  F = 0\;,
 \label{Fclosed}
\end{equation}
 and
\begin{equation}
* \;d * F = j\;.
\label{inhomMaxform}
\end{equation}
Here $*$ is the Hodge-Star operator, $d$ is the exterior derivative on forms, and $j$ denotes the current density dual vector.
Force-free electrodynamics is defined by the constraint
$$ F(j^\sharp, \chi) =0$$
for all tangent vector fields $\chi$.
Here, $\sharp$ is the unique map from the cotangent bundle to the tangent bundle $T^*({\cal M}) \rightarrow T({\cal M})$ such that $j(\chi) = g(j^\sharp, \chi)$. Note, as usual, $\sharp(j)$ is written as $j^\sharp$. The inverse of this map is denoted by $\flat$. 
The Maxwell Field tensor $F$ is said to be magnetically dominated whenever $F^2 >0$, $F$ is electrically dominated whenever $F^2 <0$, and finally a  force-free electromagnetic field $F$ is null whenever $F^2 =0$. 

The following is a summary of the basic relationship between force-free fields and 2-dimensional foliations of spacetime. For a detailed derivation of these results, see \cite{Uchida1}, \cite{Uchida2}, \cite{GT14}, and \cite{Menon_FF20}. For force-free electromagnetic fields, the kernel of the field tensor $F$, denoted by $\ker F$, is a 2-dimensional distribution that is closed under Lie brackets. This means that whenever $v,w \in \ker F$, we have that $[v,w] \in \ker F$. Such subspaces of tangent vectors are referred to as an involutive distribution. Consequently, Frobenius's theorem then implies that when a force-free $F$ exists on ${\cal M}$, the spacetime manifold can be foliated by 2-dimensional integral submanifolds of the distribution spanned by $\ker F$.
The leaves of the foliation, which are the integral submanifolds of $\ker F$,  will be denoted as $ {\cal F}_a$. A particular submanifold $ {\cal F}_a$ is referred to as a {\it field sheet}. Here $a$ belongs to some indexing set $A$. The key points here are that
$$ {\cal F}_a \cap {\cal F}_b =0 \;{\rm whenever}\;a \neq b \in A\;,\;\;\cup_{a\in A} \;{\cal F}_a ={\cal M}\;.$$
The general theory of FFE splits into three categories. Two of them, namely the magnetically and electrically dominated solutions, have very similar properties, and they differ geometrically only in detail. Nonetheless, from the point of view of the initial value problem in PDEs, it should be noted that only the magnetically dominated solutions are well posed (for example see \cite{PH13} and \cite{CR16}). The general case where non-null solutions reach the null limit can sometimes lead to divergences in the field tensor. These ideas have been explored in \cite{Menon_FF20}.
\subsection*{Non-Null FFE Theory}
In this section, we will state all the central results of \cite{Menon_2021}. Consider any foliation ${\cal F} = \{{\cal F}_a: a \in A\}$ of spacetime by 2-dimensional Lorentzian manifolds ${\cal F}_a$. In the magnetically dominated case, the metric when restricted to ${\cal F}_a$ has a Lorentz signature. We will refer to such a foliation as a 2-D {\it Lorentzian foliation} of ${\cal M}$ and denote it by ${\cal F}^{\;2L}$. The leaves of such a 2-D Lorentzian foliation are denoted by ${\cal F}_a ^{\;2L}$. For any $p \in {\cal M}$ there exists an open set $U_p$ of $p$ and an inertial frame  $(e_0, e_1, e_2, e_3)$ on $U_p$ such that the tangent space for each of the leaves of the foliation, when restricted to $U_p$, is spanned by $e_0$ and $e_1$. 
We will refer to such frames as 2-D {\it  Lorentzian foliation adapted frames} and denote it by ${\bf F}_{2L}$. Note that, here, $g(e_\mu, e_\nu) = \eta_{\mu\nu}$, where $\eta_{\mu\nu}$ are the components of the Minkowski metric.

Let $V, W$ be vector fields tangent to any ${\cal F}_a ^{\;2L}$. Then the {\it shape tensor} or the {\it  second fundamental form} $\Pi$ of ${\cal F}_a ^{\;2L}$ is defined by
$$\Pi (V, W) = (\nabla_V W)^\perp\;.$$
Here $\perp$ takes the component of the vector normal to the surface ${\cal F}_a ^{\;2L}$. The mean curvature field at any point of ${\cal F}_a ^{\;2L}$ is then defined by
$$ H=\frac{1}{2} \Big[- \Pi(e_0, e_0) + \Pi(e_1, e_1)\Big]$$
in any ${\bf F}_{2L}$. Similarly, even though the complimentary orthogonal distribution of the foliation spanned locally by $e_2$ and $e_3$ may not be an involutive distribution, we may still define a ``dual" mean curvature field by the expression
$$\tilde H=\frac{1}{2} \Big[ \Pi(e_2, e_2) + \Pi(e_3, e_3)\Big]\;. $$
We are now able to state the central theorem that will enable the search for magnetically dominated solutions \cite{Menon_2021}. The analogous result in the restricted case of stationary, axis-symmetric force-free electrodynamics in a Kerr background was previously developed in \cite{Compere:2016xwa}.

\begin{theorem}
Let ${\cal F}^{\;2L}$ be any $2$-D Lorentzian foliation of ${\cal M}$ with leaves $\{{\cal F}_a ^{\;2L}, \;a \in A\}$. Let ${\bf F}_{2L}$ be a Lorentzian frame field on a starlike open set $U_p$ about any $p \in {\cal M}$. Then, up to a constant factor in $u$, $F = u \; e_2 ^\flat \wedge e_3 ^\flat$  is the unique magnetically dominated force-free electrodynamic field in $U_p$ such that
$$\ker F|_{U_p\cap {\cal F}_a ^{\;2L}} = T({\cal F}_a ^{\;2L})$$
if and only if
\begin{equation}
    dH^\flat=-d\tilde H^\flat\;,
    \label{H+tilH}
\end{equation}
where $H (\tilde H)$ are the mean (dual) curvature field associated with the foliation. Moreover, in this case, 
\begin{equation}
 d (\ln u) = 2 (H+\tilde H)^\flat\;. 
 \label{eqforu}
\end{equation}
\label{Mdomthm}
\end{theorem}
It should be kept in mind that the theorem above generates vacuum solutions as well. Note that the resulting vacuum electromagnetic field continues to be degenerate. This inclusion of vacuum solutions in the methodology stems from the fact that when $j=0$, the resulting field is trivially ``force-free".

An analogous result applies to an electrically dominated field. The only difference is that here $(e_2, e_3)$ spans the kernel of $F$ and hence forms an involutive pair, and the force-free field is given by $F = u \; e_0 ^\flat \wedge e_1 ^\flat$. The expression for $u$, in this case, continues to be given by eq. (\ref{eqforu}).
\section{The Search for Non-Null Solutions In Kerr Geometry}
In the Boyer-Lindquist coordinate system $(t,r,\theta,\varphi)$, the Kerr metric takes the form:
$$ds^2 = g_{tt} \;dt^2 + 2 \;g_{t\varphi}\; dt\; d\varphi+ \gamma_{rr}\; dr^2 +
\gamma_{\theta \theta}\; d\theta^2 + \gamma_{\varphi \varphi}\;d\varphi^2,$$
where
$$g_{tt} =  -1 + \frac{2Mr}{\rho^2}, \;\;\; g_{t \varphi}  = \frac{-2Mr a \sin^2\theta}{\rho^2}, \gamma_{rr} = \frac{\rho^2}{\Delta}\;,\;\;\;\gamma_{\theta \theta} = \rho^2, \;\;\; \gamma_{\varphi \varphi} = \frac{\Sigma^2 \sin^2\theta}{\rho^2}\;.$$
Here,
$$ \rho^2 = r^2 + a^2 \cos^2\theta \;,\;\;\Delta = r^2 -2 M r + a^2 \;,$$
and
$$\Sigma^2 = (r^2 + a^2)^2 -\Delta \; a^2 \sin^2\theta.$$
The transformation from the Boyer-Lindquist to the Black Hole coordinates $(\bar t, r, \theta, \bar \varphi)$ is given by
$$d\Bar{t}=dt+\dfrac{r^2+a^2}{\Delta}dr\;, \;\;{\rm and}\;\;
    d\Bar{\varphi}=d\varphi+\dfrac{a}{\Delta}dr\;.$$

The transformation to White Hole coordinates $( t^*, r, \theta, \bar \varphi^*)$ is given by
$$dt^*=dt-\dfrac{r^2+a^2}{\Delta}dr\;, \;\;{\rm and}\;\;
    d\varphi^*=d\varphi-\dfrac{a}{\Delta}dr\;.$$
Following O'Neill \cite{1995gkbh}, consider the canonical inertial tetrad in the exterior Kerr geometry given by
\begin{subequations}
\begin{eqnarray}
  e_0 =\frac{1}{\sqrt{\rho^2 \Delta}} \Big[(r^2+a^2) \partial_t + a \partial_\varphi\Big]\;,  \\
  e_1 =\frac{1}{\sqrt{\rho^2 }\sin \theta}\Big[a \sin^2\theta \partial_t + \partial_\varphi\Big] \;,\\
 e_2 = \sqrt{\frac{\Delta}{\rho^2}}\; \partial_r\;,\\
 e_3 =\frac{1}{\sqrt{\rho^2 }}\partial_\theta\;.
\end{eqnarray}
\label{F2L1}
\end{subequations}
A straightforward calculation shows that
\begin{subequations}
\begin{eqnarray}
    g(\nabla_{e_0} e_0, e_3) =-\frac{a^2 \cos \theta \sin \theta}{\rho^2\sqrt{\rho^2}} \;, \\
   g(\nabla_{e_0} e_0, e_2) =\frac{a^2 \cos^2 \theta (r-M) + r (Mr-a^2)}{\rho^2\sqrt{\rho^2 \Delta }} \;,\\
    g(\nabla_{e_1} e_1, e_3) =-\frac{ \cos \theta (r^2+a^2)}{\rho^2\sqrt{\rho^2 }\sin \theta} \;,\\
         g(\nabla_{e_1} e_1, e_2) =-\frac{r}{\rho^2}\sqrt{\frac{\Delta}{\rho^2}} \;,\\
       g(\nabla_{e_2} e_2, e_1)= g(\nabla_{e_3} e_3, e_1)=0\;,
       \\ g(\nabla_{e_2} e_2, e_0) = g(\nabla_{e_3} e_3, e_0)=0 \;.   
\end{eqnarray}
\label{F2L2}
\end{subequations}
\subsection{Immediate Solutions}
Clearly, $(e_0, e_1)$ as given above, forms an involutive pair (in fact, they commute). Consider a Lorentzian foliation of ${\cal M}$ generated by the involutive distribution given by the span of $(e_0, e_1)$. Routine calculations using eqs.(\ref{F2L2}) reveals that, in this case 
$$ 2( H^\flat+\tilde H^\flat) = \frac{ \cot \theta }{\sqrt{\rho^2 }} e_3 ^\flat - \frac{(r-M)}{\sqrt{\rho^2 \Delta }} e_2 ^\flat $$
\begin{equation}
  = -\cot \theta \;d\theta-\frac{(r-M)}{\Delta} dr\;.  
  \label{lnu}
\end{equation}
The above 1-form is exact. From eq. (\ref{eqforu}), the above equation is easily integrated to obtain a final expression for $u$ given by
$$u =\frac{u_0}{\sin \theta \sqrt{\Delta}}\;,$$
where $u_0$ is an arbitrary integration constant. Since all the requirements of theorem (\ref{Mdomthm}) are satisfied we get the following vacuum degenerate field
\begin{align}
F_1= \frac{u_0}{\sin\theta \sqrt{\Delta}} e_2 ^\flat \wedge e_3 ^\flat=u_0 \frac{ \rho^2}{\sin\theta \Delta} dr \wedge d\theta\;.
\end{align}
Naturally, as expected, here $\ker F_1 = {\rm span} \{e_0, e_1\}$. 
In a similar manner, the choice of the involutive pair $(e_2, e_3)$ yields another degenerate field given by
\begin{align}
F_2 = \frac{u_0}{\sin\theta \sqrt{\Delta}} e_0 ^\flat \wedge e_1 ^\flat=u_0 \; dt \wedge d\varphi\;.\end{align}
The solutions $F_1$ and $F_2$ are Hodge-star dual of each other. As shown in \cite{Menon_2021}, vacuum, degenerate, and non-null solutions come in pairs. Our canonical tetrad yields another pair of vacuum solutions given by
\begin{align}
F_3=\frac{u_0}{\sin\theta \sqrt{\Delta}} e_0 ^\flat \wedge e_3 ^\flat =-\frac{u_0}{\sin \theta}\; d\theta \wedge (-dt + a \sin^2\theta d\varphi)\;,\end{align}
when $\ker F_3 = {\rm span} \{e_1, e_2\}$
and
\begin{align}F_4=\frac{u_0}{\sin\theta \sqrt{\Delta}} e_1 ^\flat \wedge e_2 ^\flat=\frac{u_0}{\Delta} dr \wedge ( a \;dt - (r^2+a^2)\; d\varphi)\;,\end{align}
when $\ker F_4 = {\rm span} \{e_0, e_3\}$. All the above vacuum solutions have been previously presented in \cite{Menon_2021}. These solutions are not well defined on the event horizon, and $F_1, F_2$, and $F_3$ are not defined on the symmetry axis.
\subsection{Transformed Tetrads}
Given a fixed choice of a tetrad  $(e_0, e_1, e_2, e_3)$, any other tetrad  $(\bar e_0, \bar e_1, \bar e_2, \bar e_3)$ is related to the original one by a simple, spacetime-dependent transformation given by
\begin{equation} \begin{pmatrix}
\bar e_0\\
\bar e_1\\
\bar e_2\\
\bar e_3\\
\end{pmatrix}= \Lambda(x) \begin{pmatrix} e_0\\
e_1 \\
e_2\\
e_3\\
\end{pmatrix}\;.
\label{BasicLam}
\end{equation}
Here $\Lambda(x)$ is a spacetime-dependent general homogeneous Lorentz transformation ${\cal L}$ satisfying  $\eta = \Lambda ^T\; \eta \;\Lambda\;.$ It then suffices to pick one tetrad and study its Lorentz transformations that are subject to theorem \ref{Mdomthm}.

This freedom gives us a six-parameter family of functions. For magnetically dominated solutions, without loss of generality, all that we require is that the pair $(\bar e_0, \bar e_1)$ is involutive and further that the foliation 
by integral submanifolds of the distribution satisfies eq.(\ref{H+tilH}). Similarly, for electrically dominated solutions, it is the pair $(\bar e_2, \bar e_3)$ that must form an involutive distribution, and once again, that the foliation by integral submanifolds of the distribution satisfies eq.(\ref{H+tilH}). Admittedly, this technique is not a recipe, but rather, a systematic method for a search for new solutions. This is currently the best-known technique available to seek benchmark analytic solutions for the non-null solutions describing the Blandford-Znajek mechanism and other force-free configurations. 
\subsection{Recovering a Previous Solution}
Consider a simple transformation of our canonical tetrad in the $(e_1, e_3)$ plane given by
$$ \begin{pmatrix}
\bar e_0\\
\bar e_1\\
\bar e_2\\
\bar e_3\\
\end{pmatrix}=  \begin{pmatrix} 1 & 0 & 0 & 0 \\
0 & \frac{1}{\sqrt{1+L^2}} & 0 & \frac{L}{\sqrt{1+L^2}} \\
0 & 0 & 1 & 0\\
0 & \frac{L}{\sqrt{1+L^2}} & 0 & \frac{-1}{\sqrt{1+L^2}}\\
\end{pmatrix}\begin{pmatrix} e_0\\
e_1 \\
e_2\\
e_3\\
\end{pmatrix}\;.$$
Here $L$ is an arbitrary function of the Boyer-Lindquist coordinate $r$. 
The pair $(\bar e_0, \bar e_1)$ is involutive, and here again, the foliation by the integral submanifolds satisfies eq. (\ref{H+tilH}). Remarkably, $2 (H+\tilde H)^\flat$ is again given by eq.(\ref{lnu}), and so the magnetically dominated solution, in this case, is then given by
\begin{align}
F_5= \frac{u_0}{\sin \theta \sqrt{\Delta}}\; \bar e_2 ^\flat \wedge \bar e_3 ^\flat\;.\end{align}
In Boyer-Lindquist coordinates, this becomes
\vskip0.2in

\begin{equation}
  F_5= -\frac{u_0}{\Delta} \frac{1}{\sqrt{1+L^2}} dr \wedge \left[\frac{\rho^2}{\sin \theta} d\theta + L ( a \;dt - (r^2+a^2) d\varphi)\right]\;. 
   \label{FM1sol}
\end{equation}

This is precisely the magnetically dominated solution presented in \cite{MenonTetrad15} albeit in a $3+1$ formalism of electrodynamics using electric and magnetic fields, and later in \cite{Menon_2021} as an example of theorem(\ref{Mdomthm}). The current density vector $j_5$ in this case is given by

\begin{align}
j_5&=\frac{-u_0 L^\prime}{(1+L^2)\sin \theta \sqrt{\rho^2}}\; \bar e_1 =\frac{-u_0 L^\prime}{(1+L^2)^{3/2} \rho^2 \sin^2\theta}\Big(a\sin^2\theta \partial_t + \partial_\varphi + L \sin \theta \partial_\theta\Big).\end{align}
Here $L^\prime = dL/dr$.

\subsection{New Solutions}
In this section, we show how new tetrads satisfying the requirements of theorem (\ref{Mdomthm}) may be constructed using simple general homogeneous Lorentz transformations. In the process, we have managed to construct four different sets of exact solutions in Kerr, thereby tripling the number of previously known solutions. Unfortunately, unless otherwise stated, these solutions are not defined along the symmetry axis and the event horizon of the Kerr black hole. 
\subsubsection*{I}

In the previous example, we considered a rotation in the $(e_1, e_3)$ plane. Now consider a Lorentz boost in the $(e_0, e_2)$ plane along with a  label change $e_1 \leftrightarrow e_3$ given by
$$ \begin{pmatrix}
\bar e_0\\
\bar e_1\\
\bar e_2\\
\bar e_3\\
\end{pmatrix}= \frac{1}{C}\begin{pmatrix} \sqrt{C^2+g^2} & 0 & -g & 0 \\
0 & 0 & 0 & C \\
-g & 0 & \sqrt{C^2+g^2}& 0\\
0 & C & 0 & 0\\
\end{pmatrix}\begin{pmatrix} e_0\\
e_1 \\
e_2\\
e_3\\
\end{pmatrix}\;.$$
Here $g$ is an arbitrary function of $\theta$ and $C>0$ is a constant. It is easy to check that $(\bar e_2, \bar e_3)$ forms an involutive pair, and here again, the foliations by the integral submanifolds satisfy eq. (\ref{H+tilH}). Surprisingly, once again $2 (H+\tilde H)^\flat$ is given by eq.(\ref{lnu}), and so the electrically dominated solution, in this case, is then given by
$$F_6= \frac{u_0}{\sin \theta \sqrt{\Delta}}\; \bar e_0 ^\flat \wedge \bar e_1 ^\flat\;.$$
In Boyer-Lindquist coordinates, this becomes
\vskip0.2in

\begin{equation}
 F_6=\frac{u_0}{C \sin\theta} \;d\theta \wedge \left[ \sqrt{C^2+g^2} \; ( dt - a \sin^2 \theta d\varphi) +  g\; \frac{\rho^2}{\Delta}  dr\right]\;. 
\end{equation}
The current density vector $j_6$ in this case is given by
\begin{align}
j_6&= -\frac{u_0 g^\prime}{C \sqrt{\rho^2\Delta (C^2+g^2)} \sin \theta} \;\bar e_2\nonumber \\
&= \frac{u_0 g^\prime}{C\rho^2 \sin \theta} \left[ \frac{g}{\Delta\sqrt{C^2+g^2}}\Big((r^2+a^2) \partial_t +a \partial_\varphi\Big)-\partial_r \right]. \end{align}
Here $g^\prime = dg/d\theta$. In the exterior region, $F_6$ when contracted with itself gives $F_6^2=-2\:u_0^2/\Delta\:\sin{\theta}^2$, and further
$$J_6^2=\dfrac{u_0^2\:g'^2}{(g^2+C^2)\Delta\:\rho^2\sin{\theta}^2}\;,$$
showing that the current density vector is spacelike in the exterior geometry.

It is interesting to note that when $C \rightarrow 0$, the expression for $C F_6$ reduces to the null solution presented in \cite{MD07}. In this sense, the above force-free field can be considered as an electrically dominated perturbation of the null solution presented in \cite{MD07}. 
\vskip0.2in
A minor modification of the above solution leads to a magnetically dominated solution given by
\begin{equation}
 F_7=\frac{u_0}{C\sin\theta} \;d\theta \wedge \left[ \sqrt{g^2-C^2} \; ( dt - a \sin^2 \theta d\varphi) +  g\; \frac{\rho^2}{\Delta}  dr\right]\;. 
\end{equation}
Although $F_7$ is similar in appearance to $F_6$, as a magnetically dominated solution, it is generated by a completely different foliation. Here
$$ \begin{pmatrix}
\bar e_0\\
\bar e_1\\
\bar e_2\\
\bar e_3\\
\end{pmatrix}=  \frac{1}{C}\begin{pmatrix} -g & 0 & \sqrt{g^2-C^2} & 0 \\
0 & C & 0 & 0 \\
\sqrt{g^2-C^2} & 0 & -g& 0\\
0 & 0 & 0 & C\\
\end{pmatrix}\begin{pmatrix} e_0\\
e_1 \\
e_2\\
e_3\\
\end{pmatrix}\;,$$
and $(\bar e_0, \bar e_1)$ is defines the involutive distribution. Both $F_6$ and $F_7$ are undefined on the event horizon and along the symmetry axis, and further $F_7^2=-F_6^2$. The above two solutions can be combined into one and is given by the form
$$\dfrac{\sqrt{g^2-C}}{\sin{\theta}}d\theta \wedge \left(dt-a\sin{\theta}^2\:d\varphi\right)+\dfrac{g\rho^2}{\Delta\:\sin{\theta}}dr\wedge d\theta\;.$$
In this case, $C$ can take on any real value such that $g^2-C \geq 0$. Unfortunately, this solution is undefined at the event horizon and the symmetry axis given by $\theta =0$.
\subsubsection*{II}

For constants $A,B, \alpha$ and $\beta$, let
$$g(\varphi)= \alpha \cos(A\varphi)-\beta \sin (A \varphi)\;,$$
and let
$$ \begin{pmatrix}
f_1\\
\\
 f_2\\
\end{pmatrix}=
\left(\begin{array}{cccc}
-g&-g^\prime/A
\\
\\
 g^\prime/A  & -g
\end{array}\right)
\begin{pmatrix} \cos(Bt/a)\\
\\
\sin(Bt/a)
\end{pmatrix}\;.$$
In the above equation, $\prime$ denotes the derivative with respect to $\varphi$. Now consider the pure rotation along with $- e_1 \rightarrow \bar e_3$ given by
$$ \begin{pmatrix}
\bar e_0\\
\bar e_1\\
\bar e_2\\
\bar e_3\\
\end{pmatrix}=
\left[\begin{array}{cccc}
1 & 0 & 0 & 0 
\\
 0 & \dfrac{f_2}{\sqrt{\alpha ^2+\beta ^2}} & 0 & \dfrac{f_1}{\sqrt{\alpha ^2+\beta ^2}} 
\\
 0 & -\dfrac{f_1}{\sqrt{\alpha ^2+\beta ^2}} & 0 & \dfrac{f_2}{\sqrt{\alpha ^2+\beta ^2}} 
\\
 0 & 0 & -1 & 0 
\end{array}\right]
\begin{pmatrix} e_0\\
e_1 \\
e_2\\
e_3\\
\end{pmatrix}\;.$$
Then
$(\bar e_2, \bar e_3)$ defines an involutive distribution. The mean (/dual) curvature fields associated with the foliation gives 
$$2(H^\flat +\tilde H^\flat) =\dfrac{M-r}{ \Delta}dr- \dfrac{A+\cos\theta +B \sin^2\theta}{\sin\theta}d\theta\;.$$
The above expression is an exact 1-form, and from eq.(\ref{eqforu}) we get that 
\begin{equation*}
    u= \dfrac{u_0 \exp{(B\cos{\theta})}}{\sin{\theta}\sqrt{\Delta}\left(\csc{\theta}-\cot{\theta}\right)^A}.
\end{equation*}

The electrically dominated force-free field in this case is given by
\begin{align}
    F_8 &=\dfrac{u_0 \exp{(B\cos{\theta})}}{\sin{\theta}\sqrt{\Delta}\left(\csc{\theta}-\cot{\theta}\right)^A} \;\bar e_0 ^\flat \wedge \bar e_1 ^\flat \nonumber \\
&=\dfrac{u_0 \exp{(B\cos{\theta})}}{\sqrt{\alpha ^2+\beta ^2}\; \sin{\theta}\left(\csc{\theta}-\cot{\theta}\right)^A}\left[ f_1 d\theta \wedge ( dt-a\sin^2 \theta d\varphi)- f_2 \sin \theta  dt  \wedge  d\varphi
\right].
\end{align}

The current density is given by

\begin{align}
        j_8=-\dfrac{u}{\rho^2\sqrt{\Delta}}\left(\dfrac{(A+B)a^2+Br^2}{\sqrt{\alpha^2+\beta^2}}\right)\bigg[f_1\:\sin{\theta}\left(\partial_t+\dfrac{1}{a\sin^2{\theta}}\partial_\varphi\right)-\dfrac{f_2}{a}\partial_\theta\bigg].
\end{align}
For completeness, here, $F_8^2=-2u^2$ and,
\begin{align*}
j_8^2=\dfrac{4\:f_1\:f_2\:(c_1\:a^2\cos^2{\theta}+c_3)}{a^2\Delta\rho^2\sin^4{\theta}}\;.
\end{align*}
Once again, this solution is undefined at the event horizon and the symmetry axis given by $\theta =0$.
\subsubsection*{III}
Now consider the following new tetrad $(\bar e_0,\bar e_1,\bar e_2, \bar e_3)$ generated by 
$$ \Lambda =
\left[\begin{array}{cccc}
\dfrac{(f_2 +f_1)^2}{2\sqrt{f_2 f_1} (f_1 -f_2 )} & -\dfrac{2 \sqrt{f_2 f1}}{f_1- f_2} & \dfrac{f_2 +f_1}{2 \sqrt{f_2 f_1}} & 0 
\\
 \dfrac{f_2 +f_1}{f_1 -f_2} & \dfrac{f_2 +f_1}{f_2 -f_1} & 1 & 0 
\\
 0 & 0 & 0 & 1 
\\
 \dfrac{f_1 -f_2}{2 \sqrt{f_2 f_1}} & 0 & \dfrac{f_2 +f_1}{2 \sqrt{f_2 f_1}} & 0 
\end{array}\right]
$$
in eq.(\ref{BasicLam}). Here,
$$f_1 =\alpha \exp\left(\dfrac{a c_1 t^*-a^2 c_1 \varphi^* -c_2 \varphi^*}{a}\right)$$
and
$$f_2 = \beta \exp\left(\dfrac{c_1 a^2\: \bar\varphi + c_2 \bar\varphi-c_1 a \bar t}{a}\right)\;.$$
Here star ($*$) and bar ($-$) labels represent the white hole and black hole coordinates respectively and $c_1,c_2,c_3$ and $c_4$ are real constants.
It is easily verified that $(\bar e_2, \bar e_3)$ defines an involutive distribution, and further 
$$ 2(H^\flat +\tilde H^\flat) =\dfrac{(c_1 r^2 + M -r -c_3)}{\Delta} dr- \cot \theta d\theta\;.$$
Therefore, eq.(\ref{eqforu}) gives that
$$u= u_0 \dfrac{2 \sqrt{f_2 f_1}}{\sin \theta \sqrt{\Delta}}\;.$$
So, the electrically dominated solution is given by
\begin{align}
F_9 &=u \: \bar e_0 ^\flat\: \wedge\: \bar e_1 ^\flat \nonumber \\ 
&=f_1 \left( -\dfrac{a}{\Delta} dt \wedge dr+dt \wedge d\varphi-\dfrac{r^2+a^2}{\Delta} dr \wedge d\varphi\right)
+ f_2  \left( \dfrac{a}{\Delta} dt \wedge dr + dt \wedge d\varphi+\dfrac{r^2+a^2}{\Delta} dr \wedge d\varphi\right)\;.
\end{align}
 The above solution can be written in a compact form in the mixed black hole and white hole coordinates as
 $$F_9= f_1 \:dt^* \wedge d\varphi^* + f_2 \;d\bar t \wedge d \bar\varphi\;.$$
For completeness, the current density is given by
\begin{align}
j_9= \dfrac{(c_3 + a^2 c_1 \cos^2\theta )}{\rho^2 \sin^2 \theta} \Bigg[ \dfrac{(f_1-f_2)}{a}\dfrac{r^2+a^2}{\Delta}\partial_t+\dfrac{(f_1+f_2)}{a}\partial_r+\dfrac{(f_1+f_2)}{\Delta}\partial_\varphi\Bigg]
\;,\end{align}
and
$$F_9 ^2=-\dfrac{8 f_1 f_2}{\Delta \sin^2 \theta}\;.$$
Depending on the sign of $c_1$ and $c_2$, $f_1, f_2$ can become unbounded in the remote future and past. Additionally,  $f_1, f_2$ are not well-defined as $\varphi$ sweeps a complete circle.
While $F_9$ is undefined along the symmetry axis of the Kerr black hole, modulo the discontinuity along $\varphi$, it is well defined at the future and past event horizon under a mild restriction. To see this, we write the solution in black hole coordinates. Then
\begin{align}
    F_9 = (f_1+f_2)\:d\Bar{t}\wedge d\Bar{\varphi} -\dfrac{2f_1}{\Delta} \left(a\: d\Bar{t}\wedge dr+(r^2+a^2)\: dr \wedge d\bar{\varphi}\right)\;.
    \end{align}
I.e., if $f_1/\Delta$ is well defined at the future horizon, then so is $F_9$. Since
  \begin{align}
      f_1& = \alpha \exp\left({c_1t^*-ac_1\varphi^*-\dfrac{c_2}{a}\varphi^*}\right)\nonumber\\
      & \propto \alpha \exp{\left(c_1\:\bar{t}-\Bar{\varphi}(a\:c_1+c_2)\right)} \: (r-r_+)^{\dfrac{2(a\:c_2-r_+\:c_1)}{r_+-r_{-}}}\;. \nonumber
  \end{align}
I.e., $F_9$ is well defined at the future horizon when
\begin{equation}
    \dfrac{2(a\:c_2-r_+ ^2\:c_1)}{r_+-r_{-}}\geq 1.
\end{equation}
A similar argument confirms regularity at the past horizon. Since
$$\frac{f_1 f_2}{\Delta} = \left(\frac{f_1 f_2}{\Delta^2} \right) \Delta\;,$$
when the regularity condition is met we see that $F_9$ becomes null at the horizons. This is in contrast to $F_1$ as explained in \cite{Menon_FF20}.
\subsubsection*{IV}
For any constant $k$, consider the Lorentz transformation for the remainder of this subsection.
$$ \begin{pmatrix}
\bar e_0\\
\bar e_1\\
\bar e_2\\
\bar e_3\\
\end{pmatrix}=
\left[\begin{array}{cccc}
\cosh \Theta  & 0 & \sinh \Theta & 0 
\\
 0 & \cos k  & 0 & \sin k  
\\
 0 & \sin k  & 0 & -\cos k  
\\
 \sinh \Theta  & 0 & \cosh \Theta  & 0 
\end{array}\right]
\begin{pmatrix} e_0\\
e_1 \\
e_2\\
e_3\\
\end{pmatrix}\;.$$

\subsubsection*{IV A}
When $\sin k \neq 0$, let $\Theta=\beta +\alpha  \left(\dfrac{a \cos \theta   \cos k }{\sin k }+t \right)$, $(\bar e_2, \bar e_3)$ form an involutive pair. In this case,
$$2 (H^\flat +\tilde H^\flat) =\frac{M-r-\alpha(r^2+a^2)}{\Delta}dr- \cot{\theta} d\theta\;,$$
in which case $u$ can be integrated to give
$$u = \dfrac{u_0}{\sin{\theta}} \exp{\left(\int \dfrac{M-r-\alpha(r^2+a^2)}{\Delta}dr\right)}.$$
The electrically dominated solution is this case is given by
\begin{align}
   F_{10A}&=u\: e_0^\flat \wedge e_1^\flat \nonumber\\
   &=\dfrac{u}{\sqrt{\Delta}} \bigg[a\sin{\theta}\sin{k}\sinh{\Theta}\:dt\wedge dr+\Delta \cos{k}\cosh{\Theta} dt\wedge d\theta-\Delta \sin{k}\sin{\theta}\cosh{\Theta}\: dt \wedge d\varphi \nonumber\\
   &-\rho^2 \cos{k}\sinh{\Theta}\:dr \wedge d\theta+(r^2+a^2) \sin{k}\sin{\theta}\sinh{
   \Theta}\:dr \wedge d\varphi+a\Delta\cos{k}\sin^2{\theta}\cosh{\Theta} d\theta \wedge d\varphi \bigg].
\end{align}
The current density vector $j_{10 A}$ is given by
\begin{equation}
    j_{10A}=\dfrac{a u \alpha \sin{\theta}}{\sin{k} \sqrt{\Delta} \rho^2}\bigg[ (r^2+a^2) \sinh{\Theta} \: \partial_t+\Delta \cosh{\Theta} \partial_r+a\: \sinh{\Theta} \partial_{\varphi}\bigg].
\end{equation}
Also, $F_{10A}^2=-2u^2$, and
$$j_{10A}^2=\dfrac{\alpha^2 a^2\sin^2{\theta}}{u^2\rho^2\sin^2{k}}.$$
 
\subsubsection*{IV B}
Similarly, when $\cos k \neq 0$, let $\Theta=\beta +\alpha  \left(t -\dfrac{a \cos\theta  \sin k }{\cos k }\right)$, $(\bar e_0, \bar e_1)$ form an involutive pair. The exact same expression for $u$ as above leads to the magnetically dominated solution given by
\begin{align}
    F_{10B}&=u\: e_2^\flat \wedge e_3^\flat\nonumber\\
    &\dfrac{u}{\sqrt{\Delta}} \bigg[-a \cos{k}\sin{\theta}\cosh{\Theta} dt\wedge dr+\Delta\sin{k}\sinh{\Theta} dt\wedge d\theta+\Delta \cos{k}\sin{\theta}\sinh{\Theta} dt \wedge d\varphi \nonumber\\ 
&-\rho^2 \sin{k}\cosh{\Theta} dr\wedge d\theta-(r^2+a^2)\cos{k}\sin{\theta}\cosh{\Theta} dr \wedge d\varphi
+a\Delta \sin{k}\sin^2{\theta}\sinh{\Theta} d\theta   \wedge  d\varphi \bigg].
\end{align}

The current density vector $j_{10B}$ is given by
\begin{align}
    j_{10B}=-\dfrac{a\alpha u \sin{\theta}}{\rho^2\sqrt{\Delta}\cos{\theta}}\bigg[(r^2+a^2)\cosh{\Theta}\partial_t+\Delta \sinh{\Theta}\partial_r+a\cosh{\Theta}\partial_\varphi\bigg].
\end{align}
Here $F_{10B}^2=-F_{10A}^2$, and $J_{10B}^2=-J_{10A}^2$.

\subsubsection*{IV C}
When
$$\Theta=\beta+\alpha \left(\varphi-ln(\csc\theta-\cot\theta) \cot{k}\right),$$
we find that $\bar{e}_2$ and $\bar{e}_3$ are involutive. Here,
$$2(H^\flat +\tilde H^\flat) =\dfrac{M-r-a\alpha}{\Delta} dr-\cot{\theta}\:d\theta$$
and $$u=\dfrac{u_0}{\sin{\theta}} \exp \left( \int \dfrac{M-r-a\alpha}{\Delta} dr\right)\;.$$
The electrically dominated solution is then given by
\begin{align}
   F_{10C}&=u\: e_0^\flat \wedge e_1^\flat\nonumber\\
   &=\dfrac{u}{\sqrt{\Delta}} \bigg[a\sin{\theta}\sin{k}\sinh{\Theta}\:dt\wedge dr+\Delta \cos{k}\cosh{\Theta} dt\wedge d\theta-\Delta \sin{k}\sin{\theta}\cosh{\Theta}\: dt \wedge d\varphi\nonumber\\
   &-\rho^2 \cos{k}\sinh{\Theta}\:dr \wedge d\theta+(r^2+a^2) \sin{k}\sin{\theta}\sinh{
   \Theta}\:dr \wedge d\varphi+a\Delta\cos{k}\sin^2{\theta}\cosh{\Theta} d\theta \wedge d\varphi \bigg].
\end{align}

The current density vector is given by
\begin{align}
    j_{10C}=\dfrac{\alpha \: u}{\rho^2\sqrt{\Delta}\sin{k}\sin{\theta}}\bigg[(r^2+a^2)\sinh{\Theta}\partial_t+\Delta \cosh{\Theta}\partial_r+a\sinh{\Theta}\partial_\varphi\bigg].
\end{align}
Here
$F_{10C}^2=-2\:u^2,$ and
$$j_{10C}^2=\dfrac{\alpha^2}{u^2\rho^2\sin^2{k}\sin^2{\theta}}.$$
\subsubsection*{IV D}
Finally when $$\Theta=\beta+\alpha\left(\varphi+ln(\csc{\theta}-\cot{\theta})\tan{k}\right)\;,$$
we find that $\bar{e}_0$ and $\bar{e}_1$ are involutive. Once again, the exact same expression for $u$ as above leads to the magnetically dominated solution given by
\begin{align}
    F_{10D}&=u\: e_2^\flat \wedge e_3^\flat \nonumber\\ 
   &\dfrac{u}{\sqrt{\Delta}} \bigg[-a \cos{k}\sin{\theta}\cosh{\Theta} dt\wedge dr+\Delta\sin{k}\sinh{\Theta} dt\wedge d\theta+\Delta \cos{k}\sin{\theta}\sinh{\Theta} dt \wedge d\varphi \nonumber \\
&-\rho^2 \sin{k}\cosh{\Theta} dr\wedge d\theta-(r^2+a^2)\cos{k}\sin{\theta}\cosh{\Theta} dr \wedge d\varphi
+a\Delta \sin{k}\sin^2{\theta}\sinh{\Theta} d\theta   \wedge  d\varphi \bigg] .
\end{align}
Here, the current density is given by
\begin{align}
    j_{10D}=\dfrac{u\alpha}{\rho^2\sqrt{\Delta}\cos{k}\sin{\theta}}\bigg[(r^2+a^2)\cosh{\Theta}\:\partial_t+\Delta \sinh{\Theta}\:\partial_r+a\cosh{\Theta}\:\partial_\varphi\bigg].
\end{align}

$$J_{10D}^2=-J_{10C}^2$$

\section{Conclusion}

Ever since its inception, the task of finding exact solutions to the equations of FFE has been daunting due to the unyielding nature of coupled non-linear partial differential equations. While the task of finding solutions themselves is difficult, it is even more challenging to find solutions that are regular at the event horizon and the symmetry axis. And for regular solutions, positive energy extraction from the black hole is still not guaranteed. As a result, even more than sixty years after Blandford $\&$ Znajek published their paper, an exact and physically meaningful solution to the equations of FFE has not been found that allows for energy extraction.

As opposed to looking for simplifying assumptions that render the PDEs solvable as has been done so far, we have utilized the geometric methods laid out in \cite{Menon_2021} that start with unique two-dimensional foliations that satisfy the conditions required for the existence of a non-null solution. Thus, we have found several new non-null solutions to the equations of FFE. Up until now, only two classes of exact solutions to FFE were known, only one of them being non-null. In this paper, we have introduced four new classes of non-null solutions. The search for solutions that are well defined along the horizon and the symmetry axis of the black hole, and further, a similar geometric search for null solutions described in \cite{Menon:2020hdk}, and its application are relegated to a future project.

\section*{Acknowledgements}

MM acknowledges partial support by the National Science Foundation under Grant No. PHY-2010109. RA acknowledges insightful discussions with Ryan Low about this work.

\printbibliography
\end{document}